%% file: paper.tex
\documentclass[conference,hidelinks]{IEEEtran}
\usepackage{cite}
\usepackage{balance}
\usepackage{xspace} 
\usepackage{url}
\usepackage[caption=false,font=footnotesize]{subfig}
\usepackage{paralist}
\usepackage{fixltx2e}
\usepackage[pdftex]{graphicx}  
\usepackage{amsmath,amssymb}
\usepackage{array}
\usepackage{textcomp} 
\usepackage{siunitx}
\usepackage{pgfplots}
\usepackage{xcolor}
\pgfplotsset{compat=newest,unit code/.code={\si{#1}},plot coordinates/math parser=false,grid style={lightgray}}
\usepgfplotslibrary{units}
\usetikzlibrary{patterns,angles,quotes,positioning,calc,shapes}

\graphicspath{{./img/}} 
 
\usepackage{ifthen} 
\newboolean{authnotes} 

\setboolean{authnotes}{true} 

\ifthenelse{\boolean{authnotes}}
{
\newcommand{\fm}[1]{\footnote{{\bf\color{blue} Fabian: #1}}}
\newcommand{\db}[1]{\footnote{{\bf\color{blue} Dominik: #1}}} 
\newcommand{\st}[1]{\footnote{{\bf\color{green!50!black} Sebastian: #1}}}
\newcommand{\mz}[1]{\footnote{{\bf\color{blue} Marco: #1}}}
}
{
\newcommand{\fm}[1]{}
\newcommand{\db}[1]{}
\newcommand{\st}[1]{}
\newcommand{\mz}[1]{}
}

\newcommand\figref[1]{Fig.~\ref{#1}}

\newcommand\secref[1]{Sec.~\ref{#1}}

\newcommand{\eg}{\emph{e.g.},\xspace}
\newcommand{\ie}{\emph{i.e.},\xspace}
\newcommand{\cf}{cf.\xspace}

\newcommand{\fakepar}[1]{\vspace{1mm}\noindent\textbf{#1.}}

\newcommand{\cps}{CPS\xspace}

\newcommand{\lwb}{LWB\xspace}

\newcommand{\lqr}{LQR\xspace}

\newcommand{\meter}{\ensuremath{\,\text{m}}\xspace}

\newcommand{\ms}{\ensuremath{\,\text{ms}}\xspace}

\newcommand{\kbps}{\ensuremath{\,\text{kbit/s}}\xspace}

\usepackage{fancyhdr}
\newcommand{\mytitle}{\textbf{Accepted final version.}
To appear in \textit{Proc. of the 1st Workshop on Benchmarking Cyber-Physical Networks and Systems (CPSBench), 2018}.\\
\copyright 2018 IEEE. Personal use of this material is permitted. Permission
from IEEE must be obtained for all other uses, in any current or future
media, including reprinting/republishing this material for advertising or
promotional purposes, creating new collective works, for resale or
redistribution to servers or lists, or reuse of any copyrighted component of
this work in other works.}
\begin{document}


\title{Evaluating Low-Power Wireless \\ Cyber-Physical Systems}

\author{
\IEEEauthorblockN{Dominik Baumann\IEEEauthorrefmark{2}\IEEEauthorrefmark{3}\quad\quad
Fabian Mager\IEEEauthorrefmark{1}\IEEEauthorrefmark{3}\quad\quad
Harsoveet Singh\IEEEauthorrefmark{2}\quad\quad
Marco Zimmerling\IEEEauthorrefmark{1}\quad\quad
Sebastian Trimpe\IEEEauthorrefmark{2}}
\IEEEauthorblockA{\IEEEauthorrefmark{2}Intelligent Control Systems Group, MPI for Intelligent Systems, Stuttgart/T\"ubingen, Germany}
\IEEEauthorblockA{\IEEEauthorrefmark{1}Networked Embedded Systems Group, TU Dresden, Germany}
\IEEEauthorblockA{\IEEEauthorrefmark{3}{\it Both authors contributed equally to this work.}}
\IEEEauthorblockA{\textsf{\{dominik.baumann, sebastian.trimpe\}@tuebingen.mpg.de \quad\quad \{fabian.mager, marco.zimmerling\}@tu-dresden.de}} }

\maketitle

\thispagestyle{fancy}	
\pagestyle{empty}

\input{abstract}
\input{introduction}
\input{subject}
\input{approach}
\input{caseStudy}
\input{conclusion}
\input{acknowledgements}

\balance
\bibliographystyle{IEEEtran}
\bibliography{IEEEabrv,refs}

\end{document}

%% file: abstract.tex

\begin{abstract}
Simulation tools and testbeds have been proposed to assess the performance of control designs and wireless protocols in isolation.
A cyber-physical system (\cps), however, integrates control with network elements, which must be evaluated together under real-world conditions to assess control performance, stability, and associated costs.
We present an approach to evaluate \cps relying on embedded devices and low-power wireless technology.
Using one or multiple inverted pendulums as physical system, our approach supports a spectrum of realistic \cps scenarios that impose different requirements onto the control and networking elements.
Moreover, our approach allows one to flexibly combine simulated and real pendulums, promoting adoption, scalability, repeatability, and integration with existing wireless testbed infrastructures.
A case study demonstrates implementation, execution, and measurements using the proposed evaluation approach.
\end{abstract}

%% file: introduction.tex

\section{Introduction}
\label{sec:introduction}

Modern cyber-physical systems~({\cps}) are increasingly embracing low-power embedded devices and wireless multi-hop communication to facilitate monitoring and control of physical systems at unprecedented flexibility and low cost.
It is to be expected that these wireless {\cps} will have to meet the same high dependability requirements as traditional wired \cps due to the mission- or even safety-critical nature of the applications they serve.
This motivates the need for a careful development process that is supported by a standard approach to evaluating the end-to-end performance and behavior of wireless \cps.

The end-to-end performance and behavior of a wireless \cps is determined by the interaction of multiple components.
In particular, there exists a strong mutual dependency between the controller and the wireless network.
For example, a controller that compensates for a certain fraction of packet loss over the wireless network may in return demand a shorter end-to-end communication delay to ensure closed-loop stability.
It is thus necessary to validate controller and wireless network together. 

Control toolboxes exist that consider time-varying network imperfections, including delay, jitter, and packet loss~\cite{cervin03,bauer14}.
These toolboxes help design controllers in simulation based on an abstract model of the network, but lack realism.
By contrast, sensor network testbeds can deliver detailed measurements of wireless network performance under real-world conditions~\cite{lim13,indriya}, but lack a physical system for testing control-over-wireless solutions.
Experimental results from such solutions are indeed limited and based on setups that are specific to the application or selection of hardware and software components~\cite{araujo14,bauer14}.
A standard approach to evaluate a larger class of wireless \cps from an application-level perspective is still missing.

To be effective, such a standard evaluation approach should meet at least the following four requirements:
\begin{itemize}
 \item \textbf{Suitable physical system:} The choice of physical system is crucial. It should be a well-known system, and its dynamics should match the timescales at which the control, computing, and network elements can possibly operate.
 \item \textbf{Realistic and versatile:} The approach should accommodate a variety of realistic control tasks and communication requirements in \cps, such as closing feedback loops over large distances as well as different communication patterns and traffic loads. It should provide interesting operating points that push state-of-the-art low-power wireless networking and embedded computing to its limits.
 \item \textbf{Agnostic to control and network:} Apart from a minimal, standard interface needed to sense and act on the physical system, the approach should be agnostic to the design and implementation of the control, computing, and network elements in terms of both hardware and software. The approach should be applicable to different \emph{system} solutions regardless of, for example, the type of controller, number of relays, device architecture, and wireless standard.
 \item \textbf{Promote adoption and reproducibility:} The approach should be affordable in terms of costs and effort required to adopt it, for example, by integrating the approach with an existing sensor network testbed. Moreover, it should provide means to reproduce experiments to some extent.
\end{itemize}

The main contribution of this paper is a novel evaluation approach that addresses all the above requirements for the emerging class of \cps based on embedded devices and low-power wireless communication.
As \cps integrate control and network elements, the approach evaluates them together against a physical system under real-world conditions.


The approach stipulates the use of one or multiple inverted pendulums as physical system.
This well-studied mechanical system exhibits fast dynamics relative to, for example, process engineering systems and involves relevant control challenges (\eg nonlinear, unstable, and underactuated dynamics).
When connected over a network, the ensemble of pendulums poses a number of interesting control tasks, such as local vs.\ remote stabilization, synchronization, and distributed control, with temporal dynamics ranging from ``very challenging'' to ``fairly relaxed.''
It is therefore representative of a broad class of real-world \cps, and allows for exploring and pushing the limits of what is possible with today's \cps technology.
By allowing real pendulums to be replaced by simulated ones, the proposed approach is affordable, scalable, easy to adopt, and promotes repeatability in terms of the behavior of the physical system.

After a short introduction to networked control in \secref{sec:subject}, we detail the approach in \secref{sec:approach} and describe the supported evaluation scenarios, the resulting requirements for control and communication, and the relevant performance metrics.
Using stabilization of an inverted pendulum across a multi-hop low-power wireless network as example scenario, we demonstrate in \secref{sec:caseStudy} the practical application of the proposed approach.
\secref{sec:conclusion} provides concluding remarks and a brief outlook.

%

%% file: subject.tex

\section{A Primer on Networked Control}
\label{sec:subject} 

\begin{figure}[t]
	\centering
	\includegraphics[scale=0.8]{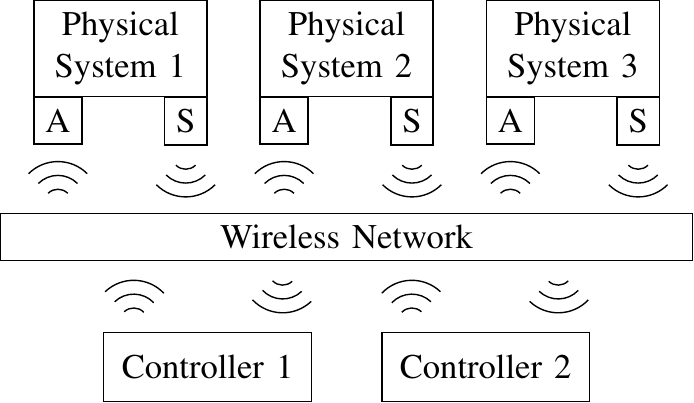}
	\caption{Typical elements of a wireless \cps.  Multiple physical systems with sensors~(S) and actuators~(A) are connected to multiple controllers via a wireless network.}
	\vspace{-5mm}
	\label{fig:components}
\end{figure}

Typical core components of a wireless \cps are depicted in \figref{fig:components}; multiple physical systems with actuators (A) and sensors (S) are connected to several controllers via a wireless network.
The sensors measure some quantity of the physical system, and their measurements are sent across the network to the controllers.  
These compute actuator commands based on the sensor data, which are again sent over the network causing the actuators to influence the system (\eg apply a force) in a certain way.
The control algorithms are designed to achieve a certain behavior of the physical system such as set-point tracking, or stabilization.
Since (multiple) feedback loops are closed over a network, such systems are also referred to as \emph{networked control systems} (NCS) \cite{HeNaXu07,lunze2014control}.

The concept of NCS is useful for a wide range of applications and different physical systems to be controlled.  However, different physical processes may pose varying requirements and challenges for the overall NCS design.  The dynamics of the physical processes play a major role: for example, processes with fast dynamics (\eg mechanical) pose more stringent requirements on network communication characteristics such as the transmission rate than systems with slow dynamics (\eg process plants); and unstable systems are more challenging than stable ones.
Closing feedback loops over a wireless network is  especially interesting in multi-agent scenarios with many systems and many controllers, for example, due to the flexibility and ease of maintenance and installation~\cite{PaJo17}. 
Examples for such wireless \cps include autonomous driving, where the vehicles  exchange information to form a platoon for fuel saving; factory automation, where the controller typically resides at some remote location and controls multiple plants; or quadcopters coordinating their flight for maintaining a specific formation.

Different control architectures and designs are used for NCS depending on, \eg the available sensors, the system size, and the control objectives.  In principle, a single controller could process data from all sensors and compute commands for all actuators.  While such a centralized controller has complete information, which is advantageous for control design,  
a centralized approach quickly becomes intractable for large-scale systems (\eg autonomous driving).
Instead, the control problem may be divided into manageable sub-problems and 
a decentralized or distributed design be employed, where actuator commands are computed on multiple controllers, \cite{Si91,Ba08}.  However, owing to the availability of only partial sensor or state information on each individual controller, distributed and decentralized designs are typically more challenging than a centralized design.  Moreover, one can distinguish problems, where the controller has access to sensor measurements of all states of the dynamic system (state-feedback controllers), versus settings with measurements of only a subset or function of the state (output-feedback control).  
While the former situation is very rare in practice, state estimation \cite{Si06} can be used to estimate the state from sensor measurements, which can then be used in combination with state-feedback control.



While NCS offer great advantages over traditional control systems with respect to, \eg installation cost, maintenance, and flexibility, the communication network does pose significant challenges for control design.
In classical control, communication is assumed to be perfect: no delay, no jitter, and no packet loss.
As long as dedicated wired communication links are considered, this assumption may not be too restrictive. 
This changes completely, however, when feedback loops are closed over wireless networks such as in most CPS application scenarios. 
For example, wireless networks are typically subject to unpredictable and often correlated packet losses and delays. 
These and other effects of imperfect communication must be taken into account in the controller design in order to ensure closed-loop stability.

%% file: approach.tex
\section{Proposed Evaluation Approach}
\label{sec:approach}

We present an approach to systematically evaluate the end-to-end performance and behavior of wireless \cps.
Motivated by the characteristics of the targeted networks and devices, this section first motivates our choice of physical system and discusses the corresponding control problems and communication requirements.
Then, we describe the supported evaluation scenarios and relevant performance metrics, and finally comment on a few practical considerations when adopting our approach.

\subsection{Characteristics of Low-Power Wireless Networks}
The proposed approach targets the class of \cps in which the wireless network (see \figref{fig:components}) consists of distributed embedded devices equipped with low-power wireless radio transceivers.
This technology brings great flexibility to the physical deployment of \cps, especially if the devices are capable of multi-hop communication and run on batteries.
Applications ranging from in-body networks to autonomous drones come into reach.

This flexibility comes at the price of resource scarcity.
State-of-the-art platforms feature 16- or 32-bit microcontrollers running at a few tens of MHz and half-duplex radios transmitting at 250--1,000\kbps.
Thus, the minimum communication delay for a single packet across a few hops is a few milliseconds.\footnote{It takes about 2\ms to transmit a 64-byte packet over one hop using ZigBee. Thus, in a 3-hop network, the minimum communication delay is about 6\ms. }
If multiple packets (\eg from sensors) need to be transmitted across the network, then effects such as medium contention, time-division multiplexing, and packet buffering will increase the communication delay.
Together with processing times (\eg to compute control inputs), the minimum end-to-end delay in low-power wireless networks is a few tens of milliseconds.


\begin{figure}[t]
	\centering
	\input{Pole-Cart.tex}
	\hspace{0.5cm}
	\includegraphics[width=0.25\linewidth]{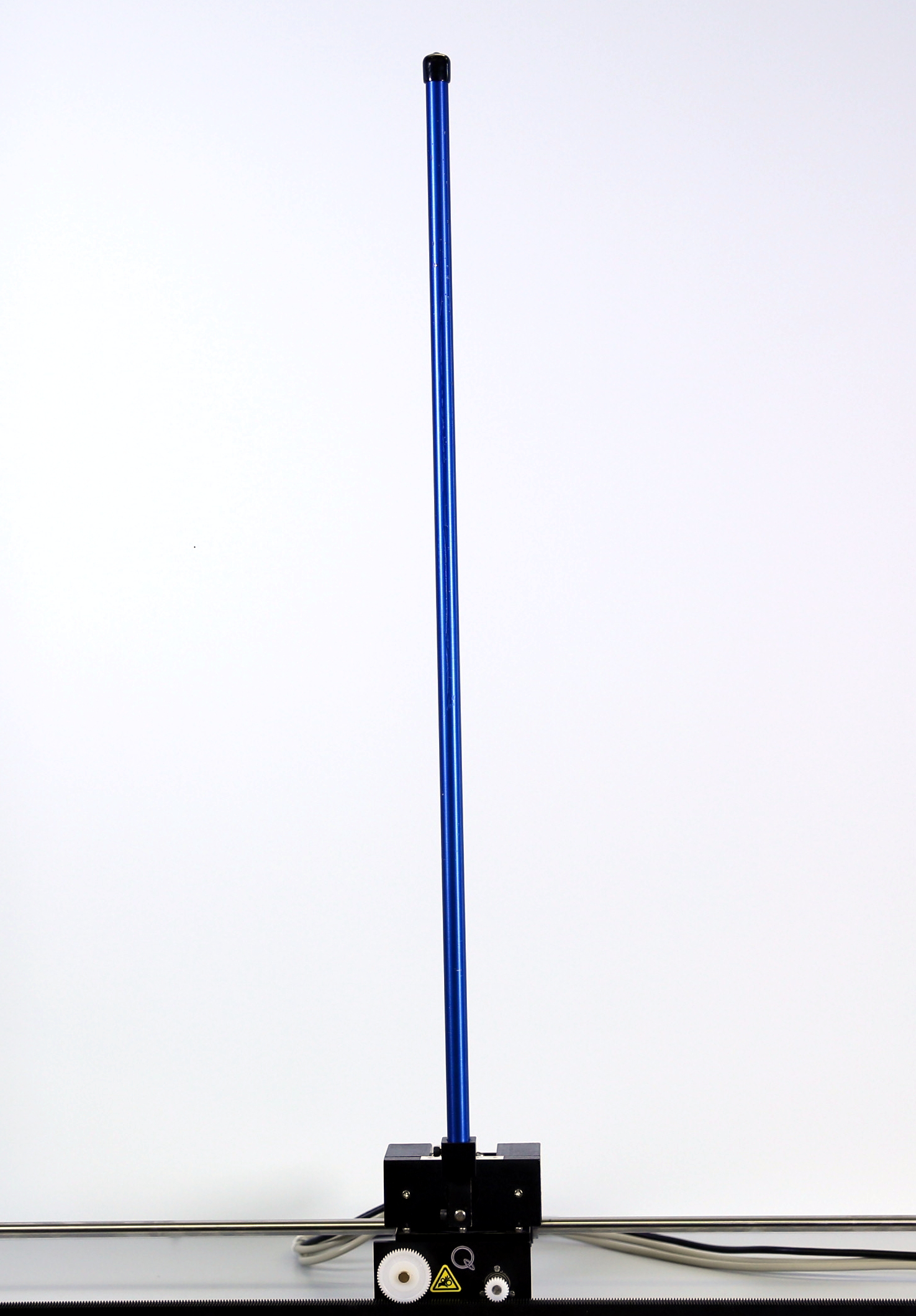}
	\caption{Left: Schematic of the inverted pendulum, right: the real system.}
	\vspace{-5mm}
	\label{fig:pendulum}
\end{figure}

\subsection{Choice of Physical System: Inverted Pendulum}
The applications for wireless \cps include a huge variety of different physical systems with diverse properties (\eg slow or fast temporal dynamics).
For a meaningful evaluation approach, a physical system is required that involves dynamics that are feasible of being controlled via a low-power, possibly multi-hop wireless network and, at the same time, allow us to push the state of the art in low-power wireless technology to its limits.  
Mechanical systems typically require update rates at the order of tens of milliseconds (or faster), which is well above, for example, typical process engineering systems such as fluid tanks \cite{araujo14}.  Thus, mechanical systems form a suitable class for evaluating and pushing low-power wireless technology for fast real-time feedback control.

A mechanical system, which is well-known and -studied, and also manageable in terms of size, affordability, and portability is the cart-pole system as shown in \figref{fig:pendulum}.  The cart-pole consists of two rigid bodies: a cart, which can move horizontally, and a pole attached to it via a revolute joint. 
The cart is actuated through a DC motor; the motor voltage is the system's input, which will cause the cart to move horizontally.  
The cart motion exerts forces on the pole; like that it is possible to influence the angle of the pendulum.
The system thus has two degrees of freedom; as only one of these is actuated, it constitutes an underactuated mechanical system.  
Cart-pole and other pendulum systems in various forms have been used extensively for research and education in control and dynamics systems (see, \eg \cite{trimpeCSM12} and references therein).

The dynamics of the cart-pole can be described as a first-order differential equation
\begin{equation}
\dot{x}(t) = f(x(t), u(t), d(t)),
\label{eq:system_dynamics}
\end{equation}
with control input $u(t)$ the motor voltage, the state $x(t)$, and a disturbance signal $d(t)$ capturing, \eg external disturbances, noise, or model mismatch.
The state $x(t)$ for the considered cart-pole  in \figref{fig:pendulum} is composed of the angle~$\theta(t)$, the cart position~$s(t)$, the angular velocity~$\dot{\theta}(t)$, and the cart velocity~$\dot{s}(t)$.
A subset of the states, namely the angle and the position, can be measured by angle sensors, whose measurements are typically corrupted by  noise or encoder quantization.
The velocity states can be estimated, \eg using finite differences or appropriate filtering \cite{Si06}.

The cart-pole system has two equilibria (\ie configurations where $\dot{x}(t)=0$): one stable equilibrium with the pendulum hanging straight down ($\theta\!=\!\SI{180}{\degree}$ in \figref{fig:pendulum}), and an unstable one with the pendulum in the upright position.  The latter is called the \emph{inverted pendulum}.

\subsection{Control Tasks}
When feedback control loops are closed over a wireless network (\cf \figref{fig:components}), network and control have a strong interaction requiring careful co-design of both parts.
Typically, controllers are designed according to the control task and the characteristics of the physical system.
On the one hand, wireless communication has to meet requirements of the closed-loop control system (\eg support required update rates), while, on the other hand, it also influences the controller design because imperfections of the wireless network (\eg delays or packet loss due to interference) have to be considered in the design. We next outline typical control tasks, which can readily be implemented in the proposed benchmarking approach and used to test the performance of wireless networking and the CPS design as a whole.

\fakepar{Stabilization}
A common control task is the stabilization of a system with unstable dynamics; stabilization of an unstable equilibrium is a typical example.
Since the inverted pendulum is inherently unstable, continuous adjustments of the cart are required to keep the pendulum upright.  This is achieved through a feedback control system that measures the system state and computes an appropriate voltage to be applied to the cart motor.
The fast dynamics of the cart-pole system require fast update rates in order to achieve  adequate performance and robustness.
In general, the higher the update rate, the better the control performance.

\fakepar{Synchronization}
Apart from stabilizing the inverted pendulum, we also consider synchronization of multiple pendulums.
To study synchronization tasks, we propose to interconnect and control multiple cart-pole systems via one low-power wireless network.  Different synchronization tasks are then conceivable.
The cart-pole system has four state variables, which could, in principle, be synchronized. 
For example, by synchronizing the cart positions, the goal is to have all carts moving in concert.
Mathematically speaking, we define the error $e_{ij}(t) = s_i(t)-s_j(t)$ for the deviation between cart $i$ and cart $j$.
By the choice of an appropriate controller, we want this error to asymptotically become zero for all cart combinations, \ie $\lim_{t \to \infty} e_{ij}(t) = 0 \quad\forall i,j.$
Since the dynamics of the individual cart-pole systems are almost identical in such a setting, it is possible to synchronize the entire state~\cite{Lu10}.

Synchronization of multi-agent systems has been studied extensively from many perspectives~\cite{Lu10,ScSe09}, for example as a consensus problem~\cite{OlSaFaMu}.
An application example of synchronization is vehicle platooning, where different vehicles try to synchronize their velocity for energy-efficient driving.
One possibility to achieve synchronization is to introduce an augmented state vector, including the states of all agents that should be synchronized.
It is then be possible to derive a state-feedback controller for the entire system (\eg through optimal control \cite{anderson2007optimal}), which can be implemented in a distributed fashion, where each agent computes its control input based on the augmented state vector.
This approach has been realized in~\cite{trimpeCSM12}, where a wired communication network was used.
While this approach allows for achieving optimal control, it  
assumes that all agents know each other's state, which requires many-to-many communication support and may increase bandwidth demands.
Alternative approaches include  
synchronization by means of nearest neighbor communication.
Under certain conditions, this results in synchronized behavior of all agents~\cite{JaLiMo03}.

\fakepar{Other}
While we primarily focus on stabilization and synchronization herein, other typical control tasks such as reference tracking or rejection of particular disturbances may also represent relevant test cases.

\subsection{Supported Test Cases and Realization}
\begin{table}[bt]
	\centering
	\caption{Design variables of the multi cart-pole benchmark system}
	\label{tab:variables}
	\begin{tabular}{|ll|}
	\hline
	Variable & Value \\ \hline
	Control task & \{stabilization, synchronization, ...\} \\ 
	No. of agents & \{1, ..., n\} \\ 
	No. of controllers & \{1, ..., m\} \\ 
	Controller location & \{local, remote\} \\
	Network extent & \{no. of nodes, no. of hops\} \\ 
	Node mobility & \{stationary, mobile\} \\ \hline
	\end{tabular}
\end{table}

Table \ref{tab:variables} gives an overview of typical variables of the design space of wireless \cps, which can be realized and tested with the proposed approach.
By selecting concrete values for each variable, a specific scenario can be derived.
As already described, we mainly consider two control tasks, stabilization and synchronization.
The number of cart-pole systems, \ie the number of control loops connected to the network, can be increased to test, \eg the bandwidth limits of the network.
This is also possible with regard to the number of controllers, \eg each cart-pole system can be controlled by a separate or by a central controller.
A further variable is the location of the controller, which can be placed locally at the pendulum or elsewhere in the network, so that sensor and actuation signals must be communicated via multiple hops.
Applications for a remote controller location can be found in factory automation, where sensor information from different systems are sent to a central controller that computes the control input and sends it back to the systems.
It is also important to specify the extent of the network in terms of number of nodes and hops, as different platforms and protocols make different compromises regarding throughput, range, energy consumption and robustness.
In addition, wireless nodes in the network can be either stationary or mobile, resulting in changing topologies and requiring protocols that support this dynamic.

\fakepar{Scenarios}
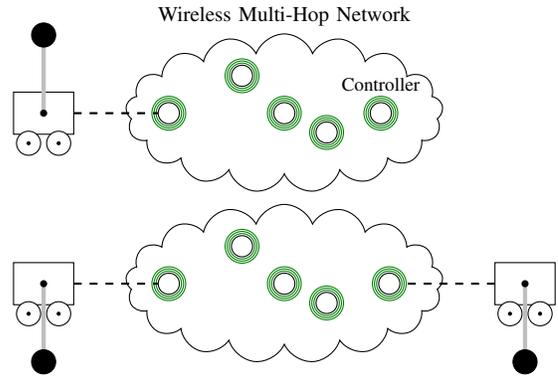
\begin{figure}[t]
	\centering
	\input{scenarios_stab.tex}
	 
	\input{scenarios_sync.tex}
	\caption{Schematics of a stabilization (top) and a synchronization (bottom) scenario. In each case, communication is over a multi-hop wireless network.}
	\vspace{-5mm}
	\label{fig:scenarios}
\end{figure}
Based on the presented design space, exemplary scenarios are described in the following.\\
\emph{Multi-hop Stabilization:} Consider one cart-pole system, one controller, and a stationary multi-hop network as depicted in the upper part of \figref{fig:scenarios}.
The controller, which is located a certain number of hops away from the physical system, stabilizes the pendulum in an upright position.\\
\emph{Multi-hop Multi-agent Synchronization:} In the lower part of \figref{fig:scenarios}, two remote cart-pole systems are synchronized with respect to their cart movements over a stationary multi-hop network.
The individual pendulums are operated around the downward equilibrium at $\theta\!=\!\SI{180}{\degree}$ so as to avoid the need for stabilization.
In this scenario, required update rates are expected to be lower than in the stabilization scenario.\\
\emph{Multi-hop Multi-agent Stabilization and Synchronization:} The combination of both previous scenarios with simultaneous stabilization and synchronization offers a challenging scenario.
It is possible to first stabilize the pendulums locally (\ie feedback loops closed locally and \emph{not} over the network) and only synchronize the controlled pendulums over the network.
The most challenging task, however, consists in doing both remote stabilization and synchronization over the network.

\fakepar{Metrics}
The evaluation of a particular scenario can look at the entire system or individual parts of it.
In the first case, we are talking about primary metrics that describe the end-to-end performance of the system, such as control quality and energy consumption.
Quality of control can be expressed in multiple ways and depends on the control task.
Examples are the quadratic error between the desired state (\eg all state variables equal to zero for stabilization) and the actual state, or the intensity of the actuation signals.
Primary metrics allow for comparing different \cps implementations of the same scenario.
Secondary metrics, on the other hand, only evaluate individual parts of the system.
For example, the network can be analyzed with regard to classical network metrics such as packet drop rate, latency, or radio duty cycles.
On the control side, the analysis could include packet drop tolerance and robustness to disturbances, for example.

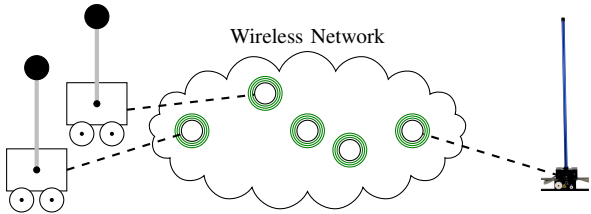
\begin{figure}
\centering
\input{scenarios_sim.tex}
\caption{Example setup with one real and two simulated pendulums connected to a wireless multi-hop network.  The flexible use of simulated and real cart-pole systems allows for easy configuration of the testbed to various benchmarking needs, for example, to investigate scalability to many systems.}
\vspace{-5mm}
\label{fig:simandreal}
\end{figure}

\fakepar{Simulated and Real Pendulums}
The proposed approach allows us to test controller and network architectures with either multiple physical pendulums, or simulated pendulums, or a combination of both.
\figref{fig:simandreal} illustrates an example setup with two simulated pendulums and one real pendulum connected to the wireless network.
The simulation includes a non-linear mathematical model of the dynamics of the pendulum \eqref{eq:system_dynamics} and is connected to a co-located node of the network, \eg via a serial interface.
It provides sensor values that must be transmitted to the controller and applies received actuation signals as it propagates the system.
A simulated pendulum increases the development speed and makes it easier to adopt our approach or integrate it into an existing infrastructure.
It helps to experiment with a larger number of pendulums to evaluate scalability properties, and in addition, traces can be recorded on real hardware and played back in the simulation to simplify testing and debugging, since the simulation offers a repeatability compared to a real system, which is always slightly different in its behavior and also subject to wearout.

A real cart-pole system is usually equipped with sensors that measure the position of the cart and the angle of the pendulum; typical examples are 
digital encoder sensors.
To read the signals from these sensors, quadrature decoders and crystals are needed.
Control inputs are connected to the cart motor with an analog voltage signal.
Depending on the specific microcontroller, this is either immediately possible or a digital to analog converter is required.
It is also useful to include a relay in the setup to prevent the cart motor from receiving random voltage signals if the microcontroller is programmed while connected to the motor.

%% file: Pole-Cart.tex
\begin{tikzpicture}[scale=0.6, every node/.style={scale=0.6}]]
\tikzstyle{ground}=[fill,pattern=north east lines,draw=none,minimum width=5,minimum height=0.1]
\draw(0.5,0.35)rectangle(-0.5,-0.35);
\draw[fill=white](-0.25,-0.5)circle(0.2);
\draw[fill=white](0.25,-0.5)circle(0.2);
\draw[ultra thick,lightgray](0,0)node(cart){}--(1,3.75);
\draw[fill=black](1,3.75)circle(0.2)node(pole){};
\draw[dashed] (0,0.5) -- (0,3.75)node(top){};
\draw[fill=black](0,0)circle(0.05);
\draw[fill=black](-0.25,-0.5)circle(0.025);
\draw[fill=black](0.25,-0.5)circle(0.025);
\node[ground,minimum width=100,anchor=north](floor)at(0,-0.7){};
\draw(floor.north east)--(floor.north west);
 \draw pic["$\theta$",draw=black, -,  angle eccentricity=1.5,angle radius =
  1cm] {angle = pole--cart--top};
  \draw[->](0,0) -- (2,0)node[below]{$s$};
\end{tikzpicture}

%% file: scenarios_stab.tex
\usetikzlibrary{shapes}

\begin{tikzpicture}[scale=0.8, every node/.style={scale=0.8}]
\tikzset{>=latex}

\node[ align = center ,draw , text = black, label = above:{Wireless Multi-Hop Network} , cloud ,cloud puffs =17 , cloud puff arc =140 ,  aspect =2.5,fill=white,minimum width = 15em,minimum height = 7.5em](nw){};
\node[draw,circle,minimum size = 10pt]at([yshift=2em,xshift = 2em]nw.south)(start1){};
\node[draw,circle,minimum size = 12pt, green!50!black]at(start1.center){};
\node[draw,circle,minimum size = 14pt, green!50!black]at(start1.center){};
\node[draw,circle,minimum size = 16pt, green!50!black]at(start1.center){};

\node[draw,circle,minimum size = 10pt]at([yshift=-2em,xshift = -2em]nw.north)(start2){};
\node[draw,circle,minimum size = 12pt, green!50!black]at(start2.center){};
\node[draw,circle,minimum size = 14pt, green!50!black]at(start2.center){};
\node[draw,circle,minimum size = 16pt, green!50!black]at(start2.center){};

\node[draw,circle,minimum size = 10pt]at([xshift=2em]nw.west)(start3){};
\node[draw,circle,minimum size = 12pt, green!50!black]at(start3.center){};
\node[draw,circle,minimum size = 14pt, green!50!black]at(start3.center){};
\node[draw,circle,minimum size = 16pt, green!50!black]at(start3.center){};

\node[draw,circle,minimum size = 10pt]at([xshift=-2.9em]nw.east)(start4){};
\node[draw,circle,minimum size = 12pt, green!50!black]at(start4.center){};
\node[draw,circle,minimum size = 14pt, green!50!black,label = above:{\small{Controller}}]at(start4.center){};
\node[draw,circle,minimum size = 16pt, green!50!black]at(start4.center){};

\node[draw,circle,minimum size = 10pt]at(nw.center)(start5){};
\node[draw,circle,minimum size = 12pt, green!50!black]at(start5.center){};
\node[draw,circle,minimum size = 14pt, green!50!black]at(start5.center){};
\node[draw,circle,minimum size = 16pt, green!50!black]at(start5.center){};




\draw(-3.5,0.35)rectangle(-4.5,-0.35);
\draw[fill=white](-3.75,-0.5)circle(0.2);
\draw[fill=white](-4.25,-0.5)circle(0.2);
\draw[fill=black](-3.75,-0.5)circle(0.025);
\draw[fill=black](-4.25,-0.5)circle(0.025);
\draw[ultra thick,lightgray](-4,0)node(cart){}--(-4,1.3);
\draw[fill=black](-4,1.3)circle(0.2)node(pole){};
\draw[fill=black](-4,0)circle(0.05);

\draw[white](3.5,0.35)rectangle(4.5,-0.35);
\draw[fill=white,white](3.75,-0.5)circle(0.2);
\draw[fill=white,white](4.25,-0.5)circle(0.2);
\draw[fill=white,white](3.75,-0.5)circle(0.025);
\draw[fill=white,white](4.25,-0.5)circle(0.025);
\draw[white](4,0)node(cart){}--(4,-1.3);
\draw[fill=white,white](4,-1.3)circle(0.2)node(pole){};
\draw[fill=white,white](4,0)circle(0.05);

\draw[dashed,thick](-3.5,0) -- (start3);

\end{tikzpicture}

%% file: scenarios_sync.tex

\begin{tikzpicture}[scale=0.8, every node/.style={scale=0.8}]
\tikzset{>=latex}

\node[ align = center ,draw , text = black, cloud ,cloud puffs =17 , cloud puff arc =140 ,  aspect =2.5,fill=white,minimum width = 15em,minimum height = 7.5em](nw){};
\node[draw,circle,minimum size = 10pt]at([yshift=2em,xshift = 2em]nw.south)(start1){};
\node[draw,circle,minimum size = 12pt, green!50!black]at(start1.center){};
\node[draw,circle,minimum size = 14pt, green!50!black]at(start1.center){};
\node[draw,circle,minimum size = 16pt, green!50!black]at(start1.center){};

\node[draw,circle,minimum size = 10pt]at([yshift=-2em,xshift = -2em]nw.north)(start2){};
\node[draw,circle,minimum size = 12pt, green!50!black]at(start2.center){};
\node[draw,circle,minimum size = 14pt, green!50!black]at(start2.center){};
\node[draw,circle,minimum size = 16pt, green!50!black]at(start2.center){};

\node[draw,circle,minimum size = 10pt]at([xshift=2em]nw.west)(start3){};
\node[draw,circle,minimum size = 12pt, green!50!black]at(start3.center){};
\node[draw,circle,minimum size = 14pt, green!50!black]at(start3.center){};
\node[draw,circle,minimum size = 16pt, green!50!black]at(start3.center){};

\node[draw,circle,minimum size = 10pt]at([xshift=-2.5em]nw.east)(start4){};
\node[draw,circle,minimum size = 12pt, green!50!black]at(start4.center){};
\node[draw,circle,minimum size = 14pt, green!50!black]at(start4.center){};
\node[draw,circle,minimum size = 16pt, green!50!black]at(start4.center){};

\node[draw,circle,minimum size = 10pt]at(nw.center)(start5){};
\node[draw,circle,minimum size = 12pt, green!50!black]at(start5.center){};
\node[draw,circle,minimum size = 14pt, green!50!black]at(start5.center){};
\node[draw,circle,minimum size = 16pt, green!50!black]at(start5.center){};




\draw(-3.5,0.35)rectangle(-4.5,-0.35);
\draw[fill=white](-3.75,-0.5)circle(0.2);
\draw[fill=white](-4.25,-0.5)circle(0.2);
\draw[fill=black](-3.75,-0.5)circle(0.025);
\draw[fill=black](-4.25,-0.5)circle(0.025);
\draw[ultra thick,lightgray](-4,0)node(cart){}--(-4,-1.3);
\draw[fill=black](-4,-1.3)circle(0.2)node(pole){};
\draw[fill=black](-4,0)circle(0.05);

\draw(3.5,0.35)rectangle(4.5,-0.35);
\draw[fill=white](3.75,-0.5)circle(0.2);
\draw[fill=white](4.25,-0.5)circle(0.2);
\draw[fill=black](3.75,-0.5)circle(0.025);
\draw[fill=black](4.25,-0.5)circle(0.025);
\draw[ultra thick,lightgray](4,0)node(cart){}--(4,-1.3);
\draw[fill=black](4,-1.3)circle(0.2)node(pole){};
\draw[fill=black](4,0)circle(0.05);

\draw[dashed,thick](-3.5,0) -- (start3);
\draw[dashed,thick](3.5,0) -- (start4);

\end{tikzpicture}

%% file: scenarios_sim.tex

\begin{tikzpicture}[scale=0.8, every node/.style={scale=0.8}]
\tikzset{>=latex}

\node[ align = center ,draw , text = black, cloud ,cloud puffs =17 , cloud puff arc =140 ,  aspect =2.5,fill=white,minimum width = 15em,minimum height = 7.5em,label=above:{Wireless Network}](nw){};
\node[draw,circle,minimum size = 10pt]at([yshift=2em,xshift = 2em]nw.south)(start1){};
\node[draw,circle,minimum size = 12pt, green!50!black]at(start1.center){};
\node[draw,circle,minimum size = 14pt, green!50!black]at(start1.center){};
\node[draw,circle,minimum size = 16pt, green!50!black]at(start1.center){};
\node[draw,circle,minimum size = 10pt]at([yshift=-2em,xshift = -2em]nw.north)(start2){};
\node[draw,circle,minimum size = 12pt, green!50!black]at(start2.center){};
\node[draw,circle,minimum size = 14pt, green!50!black]at(start2.center){};
\node[draw,circle,minimum size = 16pt, green!50!black]at(start2.center){};

\node[draw,circle,minimum size = 10pt]at([xshift=2em]nw.west)(start3){};
\node[draw,circle,minimum size = 12pt, green!50!black]at(start3.center){};
\node[draw,circle,minimum size = 14pt, green!50!black]at(start3.center){};
\node[draw,circle,minimum size = 16pt, green!50!black]at(start3.center){};

\node[draw,circle,minimum size = 10pt]at([xshift=-2.5em]nw.east)(start4){};
\node[draw,circle,minimum size = 12pt, green!50!black]at(start4.center){};
\node[draw,circle,minimum size = 14pt, green!50!black]at(start4.center){};
\node[draw,circle,minimum size = 16pt, green!50!black]at(start4.center){};

\node[draw,circle,minimum size = 10pt]at(nw.center)(start5){};
\node[draw,circle,minimum size = 12pt, green!50!black]at(start5.center){};
\node[draw,circle,minimum size = 14pt, green!50!black]at(start5.center){};
\node[draw,circle,minimum size = 16pt, green!50!black]at(start5.center){};




\draw(-4,-0.3)rectangle(-5,-1);
\draw[fill=white](-4.25,-1.15)circle(0.2);
\draw[fill=white](-4.75,-1.15)circle(0.2);
\draw[fill=black](-4.25,-1.15)circle(0.025);
\draw[fill=black](-4.75,-1.15)circle(0.025);
\draw[ultra thick,lightgray](-4.5,-0.65)node(cart){}--(-4.5,1.05);
\draw[fill=black](-4.5,1.05)circle(0.2)node(pole){};
\draw[fill=black](-4.5,-0.65)circle(0.05);

\draw(-4,0.8)rectangle(-3,0.1);
\draw[fill=white](-3.25,-0.05)circle(0.2);
\draw[fill=white](-3.75,-0.05)circle(0.2);
\draw[fill=black](-3.25,-0.05)circle(0.025);
\draw[fill=black](-3.75,-0.05)circle(0.025);
\draw[ultra thick,lightgray](-3.5,0.45)node(cart){}--(-3.5,2);
\draw[fill=black](-3.5,1.9)circle(0.2)node(pole){};
\draw[fill=black](-3.5,0.45)circle(0.05);

\node[anchor=south west,inner sep=0] at (3.25,-1){\includegraphics[scale=0.04]{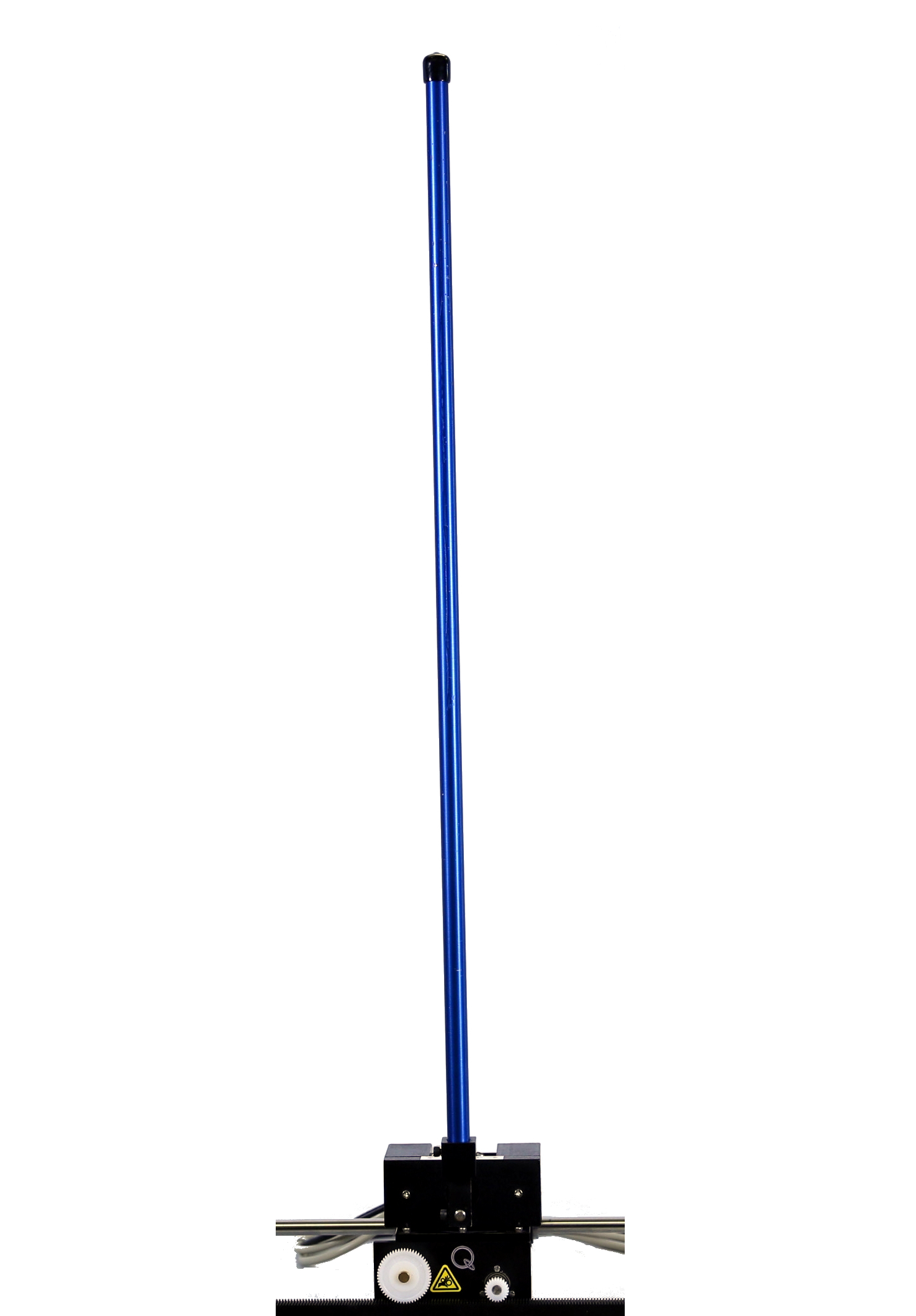}};

\draw[dashed,thick](-4,-0.65) -- (start3);
\draw[dashed,thick](-3,0.35) -- (start2);
\draw[dashed,thick](4.2,-0.75) -- (start4);

\end{tikzpicture}

%% file: caseStudy.tex

\section{Case Study: Multi-Hop Stabilization}
\label{sec:caseStudy}

\begin{figure}[t]
	\centering
	\includegraphics[width=0.8\linewidth]{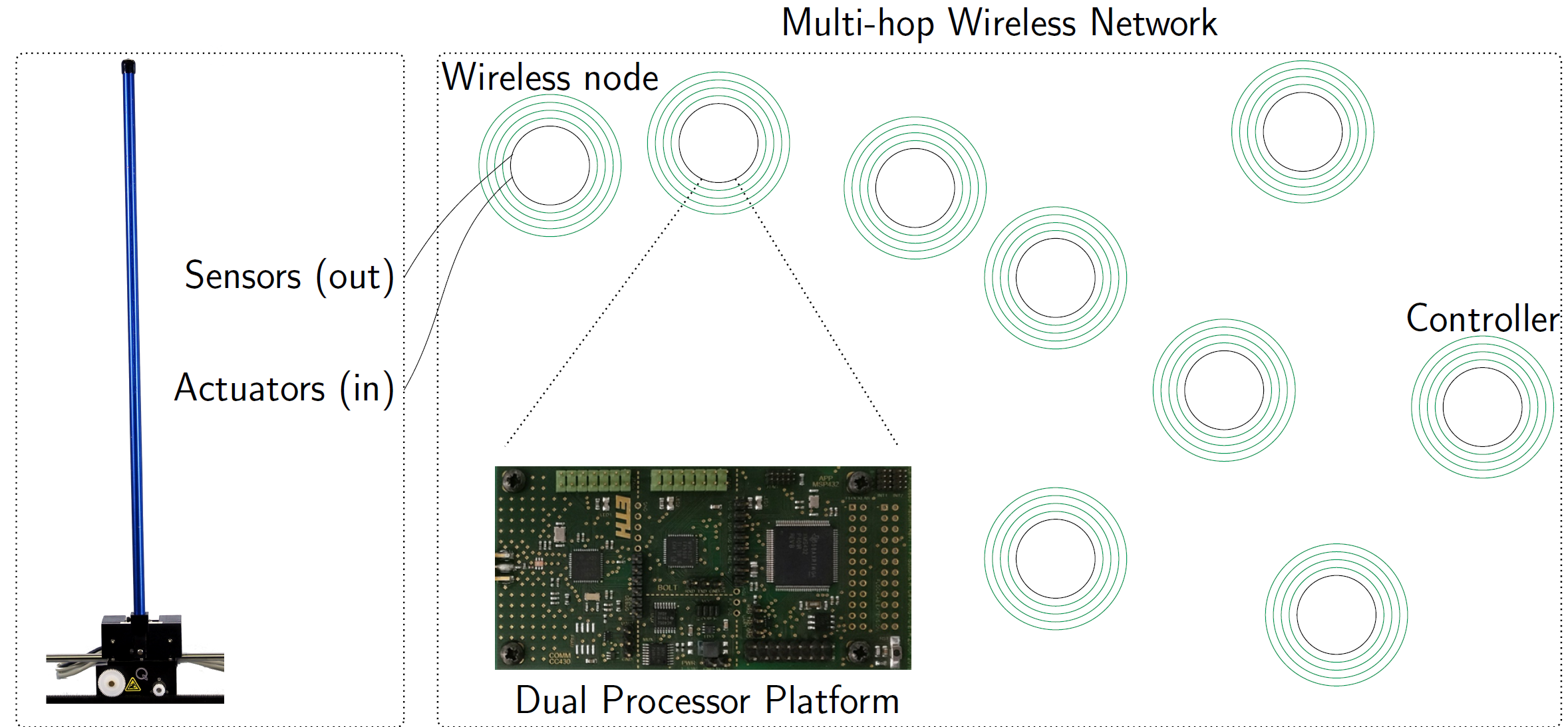}
	\caption{Testbed setup: An inverted pendulum is stabilized over a multi-hop wireless network of nine embedded devices.}
	\vspace{-5mm}
	\label{fig:testbed}
\end{figure} 

To demonstrate the feasibility of the proposed approach, an implementation of a multi-hop stabilization scenario was realized and is presented in this section as a proof of concept.
\figref{fig:testbed} shows the concrete testbed setup with a low-power wireless multi-hop network of nine embedded devices (wireless nodes) and one cart-pole system.
The system and the controller are separated from each other and located multiple hops apart.

\fakepar{Architecture}
All wireless nodes in our setup, including the controller node and the one connected to the pendulum, are dual processor platforms (using Bolt as processor interconnect~\cite{sutton15}) with off-the-shelf application (ARM-Cortex-M4, \SI{48}{\mega\hertz}) and communication processors (TI CC430, \SI{13}{\mega\hertz}).
The CC430 system on chip integrates an ultra low-power sub \SI{1}{\giga\hertz} radio transceiver running at a data rate of 250\kbps. 
We programmed the chip with a modified version of the Low-power Wireless Bus (\lwb)\cite{ferrari12}, which enables fast and energy-efficient many-to-many communication in both directions.
The application processor of the controller node executed the control algorithm.
To cope for delays and packet drops, we used state predictions based on the mathematical model provided by the manufacturer of the cart-pole system.
The predicted state is then multiplied by a gain matrix to compute a control input signal.
We computed the gain matrix by means of a linear quadratic regulator (\lqr) design~\cite{anderson2007optimal}, which represents an optimal controller for a linear system (for balancing around the equilibrium, the cart-pole dynamics are approximately linear).  This type of controller (without the wireless link) was used successfully on the platform in previous experimental studies~\cite{marco_ICRA_2017, so18}.

The application processor of the node connected to the cart-pole system was used for sensor sampling and actuation.
Sensor sampling included reading of the encoder sensors, computing the derivatives $\dot{s}$ and $\dot{\theta}$ with finite differences, and filtering measurements and derivatives with a low-pass filter.

\fakepar{Execution}
The \lwb operates with a round time of \SI{40}{\milli\second} whereby at the end of each round, the controller and pendulum have exchanged their values, \ie the controller knows the state of the pendulum, which received the control input.
This corresponds to a sample time of \SI{40}{\milli\second}.
To control the inverted pendulum, \SI{16}{\byte} of state information are sent to the controller.
The controller then computes the actuation signal (\SI{4}{\byte}) and sends it back.
Due to the strict timing requirements, the controller always computes the actuation signal based on the state information of the previous \lwb round.
The delay between sending the state information and receiving the corresponding control input is therefore twice the \lwb round time, \ie \SI{80}{\milli\second}.
In order to improve the control performance, the controller can make further predictions and send several control inputs at once, which are successively applied to the pendulum.
However, increasing the number of control inputs also increases the number of bytes that are sent.

\begin{figure}
	\centering
	\input{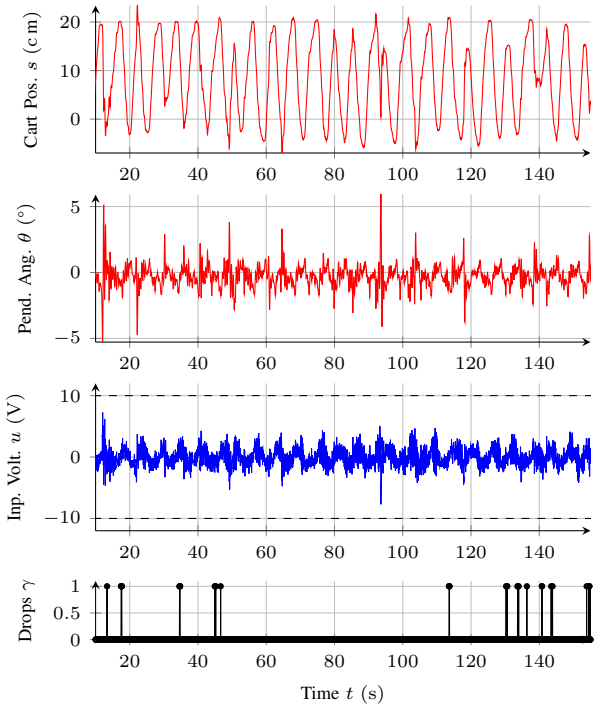}
	\caption{Results of a multi-hop stabilization experiment, showing over time, from top to bottom, the cart position $s$, the pendulum angle $\theta$, the control input $u$ with its maximum and minimum values of $\pm \SI{10}{\volt}$ (dashed lines), and the packet drops during communication $\gamma$ ($\gamma=1$ indicates a lost packet).}
	\vspace{-5mm}
	\label{fig:results}
\end{figure}

\fakepar{Results}
\figref{fig:results} shows the results of the described scenario.
Most of the time, the pendulum does not deviate more than $\theta \!=\! \SI{3}{\degree}$ from its desired position, while the cart remains within its permissible range of $\pm \SI{25}{\centi\meter}$.
In addition, the control input $u$ is within its limits of $\pm \SI{10}{\volt}$, \ie it operates in a safe regime.
From this, we conclude that the controller is able to stabilize the inverted pendulum despite occasional packet drops and an end-to-end delay of \SI{80}{\milli\second}.

%% file: conclusion.tex

\section{Conclusions}
\label{sec:conclusion}
We proposed an evaluation approach for wireless \cps based on embedded devices and low-power networking technology that meets the requirements listed in section~\ref{sec:introduction}.
Wireless \cps combine control and networking components, hence both must be evaluated together.
As an example for a concrete experimental platform, we introduced the cart-pole system, which is a mechanical system with fast dynamics that pushes low-power wireless networks to their limits when used for feedback control.
Our approach allows for different scenarios that evaluate different capabilities of the \cps regarding both network and control elements.
The possibility to include simulated pendulums facilitates adoption and integration into an existing wireless testbed infrastructure.
With a proof-of-concept implementation of multi-hop stabilization, we demonstrated the general applicability of the proposed approach.  In future work, we plan to develop and test CPS designs that tightly integrate low-power wireless communication \cite{ferrari12} and event-based control \cite{so18,Tr17} in order to enable fast, reliable, and energy-efficient control over wireless networks.

%% file: acknowledgements.tex
\section*{Acknowledgements}
This work was supported in part by the German Research Foundation (DFG) within the Cluster of Excellence
CFAED and SPP 1914 (grants ZI 1635/1-1 and TR 1433/1-1), the
Cyber Valley Initiative, and the Max Planck Society.